\documentstyle[12pt]{article}
\begin{document}

\setcounter{page}{0}
\thispagestyle{empty}

\begin{flushright}
DO -TH 94/25, TIFR/TH/94-48 \\
November 1994
\end{flushright}

\begin{center}
{\LARGE\bf  OBSERVING VIRTUAL LSPs AT LEP-II \\}
\bigskip
{\normalsize Amitava Datta
\footnote{Permanent address: Department of Physics, Jadavpur
University, Calcutta 700 032, India.}\\}
{\footnotesize Institut F\"ur Physik, Universit\"at Dortmund, D
44221 Dortmund, Germany.} \\
{\footnotesize\it Electronic
address: datta@het.physik.uni-dortmund.de \\}

{\normalsize Aseshkrishna Datta \\}
{\footnotesize Department of Physics, Jadavpur University,
Calcutta 700 032, India.} \\

{\normalsize Sreerup Raychaudhuri \\}
{\footnotesize Theoretical Physics Group, Tata Institute of
Fundamental Research, \\ Homi Bhabha Road, Bombay 400 005,
India.} \\
{\footnotesize\it Electronic address: sreerup@theory.tifr.res.in
\\}
\vskip 5pt

{\large\bf ABSTRACT}
\end{center}
\normalsize\sl

In the minimal supersymmetric extension of the Standard Model
(MSSM) with $R$-parity conservation the possibility that the
sneutrino and the second lightest neutralino decay dominantly
into invisible channels and, in addition to the lightest
neutralino, carry missing energy, is still open --- both
theoretically and experimentally.  It is shown that for a large
region of the parameter space allowed by {\rm LEP-I} data, these
particles can contribute significantly to the process $e^+ e^-
\longrightarrow \gamma$ + {\rm nothing}.  This signal, if it
exists, can be easily detected at {\rm LEP-II} as an enhancement
over the Standard Model or the conventional MSSM predictions.

\newpage
\normalsize\rm

The search for supersymmetry (SUSY) at the TeV scale is a
high-priority programme of current high energy physics. The
parameter space of even the minimal supersymmetric extension of
the standard model (MSSM)\cite{1} is, however, rather
complicated. This has not yet been sufficiently restricted by
experiments to make very accurate predictions. It should,
therefore, be emphasised that until conclusive experimental
evidence prefers one region of the above space over the others,
all analyses of experimental data should be carried out keeping
the consequences of the parameter space in mind.

As an example, let us note that the usual strategy for hunting
for superparticles hinges on one crucial assumption: that, by
virtue of a conserved quantum number --- $R$-parity --- there is
a single, stable, weakly-interacting neutral superparticle, the
so-called lightest supersymmetric particle (LSP). This particle,
if produced, easily escapes detection and carries missing
transverse energy ($\not \!\! E_T$) --- traditionally regarded
as the most distinctive signature of SUSY.  Moreover as a result
of the above  conservation law all other superparticles decay
into the LSP either directly or through cascades.  The
predictions for the latter decays are often dependent on the
parameter space.  Thus, apparently clean limits on the squark
and gluino masses derived under the assumption that the gluino
decays directly into the photino with 100 \% branching ratio
\cite{2,3} become somewhat more involved given the realistic
possibility that the gluino can decay into other channels as
well \cite{4}.  This has been recognised in the latest
experimental searches and SUSY parameter-dependent mass limits
have been published in the literature \cite{3}.

In all the experimental searches carried out to date, however,
it is still assumed that the LSP {\it alone} carries missing
transverse energy ($\not \!\! E_T$). The purpose of this note is
to draw attention to a significant region of the parameter space
of MSSM ( with $R$-parity conservation) which violates this
assumption and to suggest a particularly sensetive probe of this
hitherto unexplored space.

The minimal supersymmetric extension of the standard model
(MSSM) contains four spin-$\frac{1}{2}$ neutral particles. These
particles are the superpartners of the photon, the $Z$-boson and
the two neutral $CP$-even Higgs bosons. Linear combinations of
these four states, the four neutralinos ($\widetilde{N_i}$,
i=1,4), are the physical states.  In the currently-favoured
models, the lightest neutralino ($\widetilde{N_1}$) is assumed
to be the LSP \cite{1}.

Recently the possibility has been emphasised \cite{5,6} that
there may exist other superparticles which, though unstable,
{\it decay dominantly into invisible channels}. This can occur
if the sneutrinos ($\widetilde{\nu}$) (the superpartners of the
neutrinos), though heavier than the LSP, are lighter than the
lighter chargino ($\widetilde{\chi_2}^\pm$) and the second
lightest neutralino ($\widetilde{N_2}$) and are also lighter
than all other superparticles. As a consequence, the invisible
two-body decay mode $\widetilde{\nu} \longrightarrow \nu
\widetilde{N_1}$ opens up and completely dominates over others,
being the only kinematically-allowed two-body decay channel for
the sneutrinos. The other neccesary condition for this scheme to
work is that the $\widetilde{N_1}$ has a substantial Zino
(superpartner of the $Z$-boson) component.  This, however, is
almost always the case as long as the gluino (the superpartner
of the gluon) has a mass ($m_{\widetilde{g}}$) in the range
interesting for the SUSY searches at the Tevatron \cite{7}.
Moreover, in such cases the $\widetilde{N_2}$ --- which also has
a dominant Zino component --- decays primarily through the
process $\widetilde{N_2} \longrightarrow \nu \widetilde{\nu}$.
These two particles ($\widetilde{N_2}$ and $\widetilde{\nu}$),
decaying primarily into invisible channels, may act as
additional sources of $\not \!\! E_T$ and can significantly
affect strategies for SUSY searches \cite{5,6}.  They are,
therefore, called {\it virtual} LSPs or VLSPs in the subsequent
discussion. The above scenario, which is certainly consistent
with all the experimental results on SUSY searches, can also be
easily accommodated in the more constrained and
theoretically-motivated models based on $N = 1$ Supergravity
with common scalar and gaugino masses at a high scale
\cite{6,8}.

Some consequences of the VLSP scenario (as opposed to the
conventional MSSM where the LSP is the only source of missing
$E_T$) mainly in the context of hadron colliders  have been
discussed in refs. \cite{5,6,8,9}.  In this work, we consider
the much cleaner process $e^+ e^- \longrightarrow \gamma +
nothing (\not \!\!\! E_T)$ which will be of considerable
importance at LEP-II and any other future high energy $e^+ e^-$
colliders.  In the standard model (SM) only  $\nu
\overline{\nu}$ pairs contribute to the final state. In the
conventional MSSM both $\nu \overline{\nu}$ and
$\widetilde{N_1}\widetilde{N_1}$ pairs contribute to this kind
of effect.  With VLSPs, however, there will be additional
contributions from $\widetilde{\nu}\widetilde{\nu}$ and
$\widetilde{N_i}\widetilde{N_j}~(i, j = 1,2)$ which tend to
increase the cross-section quite significantly.  We have made a
detailed analyis of this effect and find that a 15--20 \%
enhancement of the cross-section over the prediction of the SM
occurs in a significant region of the MSSM parameter space
allowed by the experimental data (most notably from  LEP-I
\cite{10}).  Moreover, the bulk of the extra contribution comes
from $\widetilde{\nu}\widetilde{\nu}$ pairs. Thus, such a
signal, if detected, can be distinguished not only from the SM
but also from the conventional MSSM without VLSPs.

We now turn to the cross-sections relevant for the process $e^+
e^- \longrightarrow \gamma + nothing (\not \!\! E_T)$. The most
important contribution to this comes from $e^+ e^-
\longrightarrow \gamma \widetilde{\nu} \widetilde{\nu}$. The
amplitudes for the relevant Feynman diagrams are given in, for
example, ref.\cite{11} in the limit when the chargino is purely
a Wino (superpartner of the $W$-boson).  We have computed the
full cross-section taking into account the chargino-mixing
matrix. The formulae are somewhat cumbersome and will be
presented elsewhere \cite{12}.  Our numerical results, however,
agree with those given in ref. \cite{11} in the appropriate
limits. We have also computed the cross-section for the process
$e^+ e^- \longrightarrow \gamma \widetilde{N_i}
\widetilde{N_j}$, $(i,j = 1,2)$ taking the $4\times 4$
neutralino mass matrix into account (the details will be
presented in ref. \cite{12}).  Our results agree, in the
appropriate limit, with the those of ref. \cite{13} where $e^+
e^- \longrightarrow \gamma \widetilde{N_1} \widetilde{N_1}$ was
obtained in the limit when $\widetilde{N_1}$ is a photino
without any mixings. Finally we have calculated the full
cross-section for the purely SM process $e^+ e^- \longrightarrow
\gamma \nu \bar{\nu}$ and checked that it agrees with the
results of ref. \cite{14} in the appropriate limits.  It may be
noted at this point that we have considered lowest order
cross-sections only, radiative corrections being neglected. The
effects of these will be considered elsewhere \cite{12}. In any
case, the only radiative corrections likely to cause changes
which are at all significant are those due to emission of soft
photons from the initial states. This is a well-known
\cite{yellowbook} effect and is likely to shift the peak in the
photon energy distribution by a few GeV. This would not make it
necessary to change the kinematical cuts (see below) very much,
if at all.

Current bounds from LEP-I tell us that it is not possible to
produce sneutrinos from the on-shell decay of a $Z$-boson
produced in $e^+ e^-$ collisions with or without a radiated
photon. However, this is certainly possible with neutrinos. As a
result, the energy distribution of a radiated photon accompanied
by neutrinos will have a resonant peak at some value close to
$\sqrt s - m_Z$ unlike the case for accompanying sneutrinos.
This can help us reduce the signal-to-background ratio. In Fig.
1 we present the energy ($E_\gamma$) distribution of the photon
from the signal ($e^+ e^- \longrightarrow \gamma \widetilde \nu
\widetilde \nu$) for $m_{\widetilde \nu} = 50 ~{\rm GeV}, \tan
\beta = 2$ and $m_{\widetilde g} = 200,300$ GeV (upper and lower
of the dotted histograms respectively) as well as the SM
background (solid histogram) at $\sqrt s = 190$ GeV, where we
have used a cut $E_\gamma > 5$ GeV, which is dictated by
detector design and the removal of other backgrounds
\cite{sunanda}. One observes that the background has a resonant
peak in the vicinity of 75 GeV which is more or less in the
right ballpark for $\sqrt s = 190$ GeV. It is now obvious that a
cut of 5 GeV $< E_\gamma < 60$ GeV removes the peak and thereby
reduces the background by about a third without affecting the
signal significantly.

We next consider the angular distribution of the photon in the
signal and the background, where $\theta_\gamma$ is the angle
made by the photon with the beam direction\footnote{One should
note that the considerations stated above \cite{sunanda} already
impose a cut of $40^\circ < \theta_\gamma < 140^\circ$}.  We
find that without the cut on $E_\gamma$ the distributions look
very similar. With the cut, however, most of the signal is
contained in the region $40^\circ < \theta_\gamma < 120^\circ$
while the background is more or less uniformly distributed over
the entire range of consideration. This is illustrated in Fig. 2
where the conventions are identical with those of Fig. 1.
Accordingly we have chosen our second kinematical cut to be
$40^\circ < \theta_\gamma < 120^\circ$.  A combination of the
two kinematic cuts optimises the signal-to-background ratio.

In Fig. 3 we plot the combined cross-section for the processes
$e^+ e^- \longrightarrow  \gamma \widetilde{\nu}
\widetilde{\nu}$ and $e^+ e^- \longrightarrow  \gamma \nu \bar
\nu$ as a function of $m_{\widetilde{\nu}}$  at $\sqrt s = 190$
GeV. The coupling of the sneutrinos and the lighter chargino
($\widetilde{\chi_2}^\pm$) as well as the
$\widetilde{\chi_2}^\pm$ mass which are relevant for the
t-channel $\widetilde{\chi_2}^\pm$ exchange diagrams (see, {\it
e.g}, ref.\cite{11}) depend \cite{1} on the SUSY parameters
$\mu, \tan \beta$ and the gluino mass $m_{\widetilde{g}}$. We
have illustrated results for $m_{\widetilde{g}}$ = 200 GeV (the
upper dashed band in Fig. 3) and 300 GeV (the lower dashed band
in Fig 3).  The widths of these bands are due to varying $\mu$
and $\tan \beta$ over their LEP-allowed values \cite{10}
consistent with the mass-spectrum required for the VLSP
scenario.  It should be noted that the lower value of
$m_{\widetilde{g}}$ is well within the striking range of direct
searches at the Tevatron collider while the upper one is very
likely to be beyond this range. For comparison we have also
presented the cross-section for the purely SM process which
corresponds to the middle one of the three solid lines in Fig 3.
The other two solid lines are representative of the statistical
fluctuations of the number of events expected from the standard
model process assuming an integrated luminosity of 500
pb$^{-1}$.  Thus, if the cross-section in the VLSP scenario is
above the uppermost solid line, the effect cannot be
interpretated as a fluctuation.

We note from Fig. 3 that  a reasonable range of sneutrino masses
can be probed at LEP-II particularly for the relatively low
$m_{\widetilde{g}}$ case. It is also interesting to note that
even if $m_{\widetilde{g}}$ is beyond the reach of direct
searches at the Tevatron collider, it is not completely hopeless
to look for SUSY signals at LEP-II if the VLSPs are present.
This is especially so since the additional contributions of the
$\widetilde{N_i} \widetilde{N_j}$ pairs to the final state,
though rather modest by themselves (see below), push up the
signal and bring a significant part of the lower band above the
upper line.  We conclude, then, that the VLSP scenario has the
potential of enhancing the cross-section for $e^+ e^-
\longrightarrow  \gamma + nothing$ by 15 - 20 \% above the SM
prediction. We have compared this enhancement with the one that
may arise due to the addition of one more stable neutrino (with
mass comparable to the sneutrino) to the SM. We find that indeed
the enhancement due to the VLSPs is in general  comparable to
and quite often significantly larger(depending on the SUSY
parameters ) than the similar effect produced by the above
neutrino.  This effect should, in fact, be discernable at any
$e^+ e^-$ machine capable of revealing effects at the level of a
few percent.

We now turn to the neutralino contributions to the signal. In
Fig. 4 we present combined cross-sections for the processes $e^+
e^- \longrightarrow  \gamma \widetilde{N_1} \widetilde{N_1}$ and
$e^+ e^- \longrightarrow  \gamma \nu \bar \nu$ as a function of
$m_{\widetilde{l}}$ (the slepton mass) at $\sqrt s = 190$ GeV.
The two bands arise due to reasons explained above. Although we
have varied the slepton mass over the {\sl entire} LEP-allowed
range, only the values in the higher side of the range are
relevant for the VLSP scenario (since otherwise the
lepton-slepton two-body decay channel will be open to the
VLSPs).  Nevertheless, we have considered the above range for
$m_{\widetilde{l}}$ to demonstrate that in the conventional MSSM
where only LSP pairs contribute to the final state, the signal
is rather too small to be detected above the fluctuations.
Among the neutralino pairs the contributions of $\widetilde{N_1}
\widetilde{N_2}$ and $\widetilde{N_2} \widetilde{N_2}$ pairs are
even  smaller. As discussed above, all $\widetilde{N_i}
\widetilde{N_j}$ pairs taken together ($i,j = 1,2$), can,
however, enhance the signal by 5-8 \% of the SM prediction.

A distinct enhancement over the SM prediction for, $e^+ e^-
\longrightarrow \gamma + nothing$, if detected at LEP-II, cannot,
therefore, be explained by the conventional MSSM where the LSP
is the only source of missing energy. Such an observation will
strongly favour  the VLSP scenario, which essentially means that
there is a relatively light sneutrino and the parameter space of
the MSSM is quite severely restricted to the area which leads to
this scenario. On the other hand the absence of this signal will
eliminate a considerable segment of a hitherto unexplored region
of the MSSM parameter space.

{\it Acknowledgements} : The authors would like to thank D. K.
Ghosh, M. Guchait and R. Shrivastava for discussions and
software support. Amitava Datta thanks Prof E. A. Paschos for
hospitality at the University of Dortmund, and the Alexander von
Humboldt Stiftung, Bonn, for financial support and the
Department of Science and Technology, Government of India for
the grant of a research project under which part of this work
was done.  Aseshkrishna Datta thanks the Theoretical Physics
Group, Tata Institute of Fundamental Research, Bombay for
hospitality and the Council of Scientific and Industrial
Research, India for financial support. The work of SR is
supported by a project (DO No. SR/SY/P-08/92) of the Department
of Science and Technology, Government of India.

\newpage
{\centerline\Large\bf  Figure Captions}
\vskip 10pt

{\sl Fig. 1.} ~ ~The differential cross-section for the process
$e^+ e^- \longrightarrow \gamma + nothing$ as a function of
$E_\gamma$ ( $E_\gamma > 5$ GeV).  The solid histogram is the
Standard Model result~. The dotted histograms represent
contributions  due to associated production of sneutrino pairs
in the VLSP scenario for $m_{\widetilde g} =$ 200 GeV (upper
histogram) and 300 GeV (lower histogram), with $m_{\widetilde
\nu} = 50$ GeV,$\mu = - 200$ GeV, and $\tan \beta = 2$.  The
behaviour remains unchanged for different values of
$m_{\widetilde \nu},
\mu~~ and \tan \beta$.

{\sl Fig. 2.} ~ ~The differential cross-section for the process
$e^+ e^- \longrightarrow \gamma + nothing$ as a function of
$\theta_\gamma$.  Cuts of 5 GeV $< E_\gamma < 60$ GeV and
$40^\circ < \theta_\gamma < 140^\circ$ have been imposed. The
convention for the histograms and SUSY parameters are as in Fig.
1. The behaviour remains unchanged for different values of
$m_{\widetilde \nu}, \mu~~ and \tan \beta$.

{\sl Fig. 3.} ~ ~Variation of the total cross-section for the
process $e^+ e^- \longrightarrow \gamma + nothing$ as a function
of $m_{\widetilde \nu}$. The middle one of the three solid lines
indicates the Standard Model contribution , while the upper and
lower ones represent fluctuations calculated on the basis of an
integrated luminosity of 500 pb$^{-1}$. The upper (lower) dashed
band represents the enhancement due to associated production of
sneutrino pairs for $m_{\widetilde g}=$ 200 (300) GeV.

{\sl Fig. 4.} ~ ~Variation of the total cross-section for the
process $e^+ e^- \longrightarrow \gamma + nothing$ as a function
of $m_{\widetilde \l}$. The solid lines are the same as in Fig.
3.  The upper (lower) dashed band represents the added
contribution due to associated production of LSP pairs for
$m_{\widetilde g}=$ 200 (300) GeV.


\newpage
\vskip 20pt
\thebibliography{19}
\bibitem{1} For  reviews see, for example, H.P.Nilles, {\it
Phys. Rep.} {\bf 110}, 1 (1984); P.Nath, R.Arnowitt and
A.Chamseddine, {\it Applied N = 1 Supergravity}, ICTP Series in
Theo. Phys., Vol I, World Scientific (1984); H. Haber and G.
Kane, {\it Phys. Rep.} {\bf 117}, 75 (1985); S.P.Misra, {\it
Introduction to Supersymmetry and Supergravity}, Wiley Eastern,
New Delhi (1992).
\bibitem{2} UA1 collaboration C.Albajar {\it et al} {\it Phys.
Lett.} {\bf198 B} (1987) 261; UA2 collaboration J.Alitti {\it et
al} {\it Phys. Lett.} {\bf235 B} (1990) 363.
\bibitem{3} CDF collaboration, F. Abe {\it et al.}, {\it Phys.
Rev. Lett.} {\bf 69}, (1992) 3439 .
\bibitem{4} H.Baer, X.Tata and J.Woodside, {\it Phys. Rev.} {\bf
D44} (1991) 207.
\bibitem{5} A.Datta, B.Mukhopadhyaya and M. Guchhait, PREPRINT
MRI-PHY-11/93, submitted for publication to {\it Mod. Phys.
Lett.}
\bibitem{6} Some apsects of VLSPs in the context of SUSY
searches at hadron colliders have  been considered by H. Baer,
C. Kao and X. Tata, {\it Phys. Rev.} {\bf D48}, R2978 (1993); M.
Barnett, J. Gunion and H. Haber, LBL preprint LBL-34106 (1993).
\bibitem{7} See, e.g., M.Guchait, {\it Z. Phys.} {\bf
C57},(1993) 157 .
\bibitem{8} A.Datta, M.Drees and M.Guchait ,paper submitted to the
XXVII International Conf on High Energy Physics,
Glasgow,1994,Contribution Code: gls 0712 and Dortmund University
pre-print (in preparation).
\bibitem{9} A.Datta, S.Chakraborty and M.Guchait ICTP, Trieste preprint
IC/94/303 (hep-ph 9410331)
\bibitem{10} H. Baer, M Drees, X. Tata, {\it Phys. Rev.} {\bf D41},
 3414 (1990); G. Bhattacharyya, A. Datta, S. N. Ganguli and A.
Raychaudhuri, {\it Phys. Rev. Lett.} {\bf 64}, 2870 (1990); A.
Datta, M. Guchhait and A. Raychaudhuri, {\it Z. Phys.} {\bf
C54}, 513 (1992); J. Ellis, G. Ridolfi and F. Zwirner, {\it
Phys. Lett.} {\bf B237}, 923 (1990); M. Davier in Proc. Joint
International Lepton-Photon and Europhysics Conference in High
Energy Physics, Geneva, 1992 (eds.  S. Hegarty {\it et al.},
World Scientific, 1992) p151.
\bibitem{11} M.Chen et al, Phys. Rep. {\bf 159} (1988) 201.
\bibitem{12} A.Datta, Asesh Datta and S.Raychaudhuri, {\it in preparation}.
\bibitem{13} K.Grassie and P.N.Pandita, {\it Phys. Rev.} {\bf D30} (1984) 22.
\bibitem{14} K.J.F.Gaemers, R.Gastmans and F.M.Renard, {\it Phys. Rev.}
{\bf D19} (1979) 1605; ~M.Caffo, R.Gatto and E.Remiddi, {\it
Nucl. Phys.} {\bf B286} (1986) 293.
\bibitem{yellowbook} For a review and further references see, for example,
L.Trentadue, {\it  Z Physics at LEP 1}, Vol. 1 (eds. G.Altarelli
{\it et al}) p129.
\bibitem{sunanda} S.Banerjee, L3 Collaboration, {\it private communication}.
\end{document}